\documentclass[11pt,a4paper]{article}

\usepackage[T1]{fontenc}
\usepackage[utf8]{inputenc}
\usepackage{lmodern}
\usepackage{microtype}
\usepackage[margin=2.3cm]{geometry}
\usepackage{amsmath,amssymb}
\usepackage{graphicx}
\usepackage[font=small,labelfont=bf,labelsep=period,justification=justified]{caption}
\usepackage[super,sort&compress,comma]{natbib}
\usepackage{xcolor}
\usepackage[colorlinks=true,linkcolor=blue!50!black,citecolor=blue!50!black,urlcolor=blue!50!black]{hyperref}
\usepackage{setspace}
\usepackage{float}
\usepackage{booktabs}

\graphicspath{{Figures/}}

\newcommand{\LMPO}{\ensuremath{\mathrm{Li}_x\mathrm{MnPO}_4}}
\newcommand{\Mntwo}{\ensuremath{\mathrm{Mn}^{2+}}}
\newcommand{\Mnthree}{\ensuremath{\mathrm{Mn}^{3+}}}

\newcommand{\UV}{DFT\,+\,$U$\,+\,$V$}
\newcommand{\MnO}{\ensuremath{\mathrm{MnO}_6}}
\newcommand{\tdw}{\ensuremath{t_{\mathrm{dwell}}}}

\renewcommand{\figurename}{Fig.}
\newcommand{\rev}[1]{#1}
\newcommand{\revLi}[1]{#1}
\newcommand{\revviii}[1]{#1}
\newcommand{\red}[1]{#1}

\title{\textbf{Oxygens breathe, electrons jump}}

\author{Cristiano Malica$^{1,*}$ \rev{and} Nicola Marzari$^{1,2}$}

\date{\small
$^1$U Bremen Excellence Chair, Bremen Center for Computational Materials Science, and MAPEX Center for Materials and Processes, University of Bremen, Bremen, Germany.\\
$^2$\rev{Theory of Condensed Matter,} Cavendish Laboratory, Department of Physics, University of Cambridge, Cambridge, UK.\\[2pt]
$^*$e-mail: cmalica@uni-bremen.de}

\begin{document}
\maketitle

% =====================================================================
\begin{abstract}
\noindent
Electronic transport in polyanionic battery cathodes proceeds through small-polaron hopping between transition-metal sites and is commonly assumed to be tightly coupled to the motion of charge-compensating mobile ions. Here, using olivine \LMPO{} as a prototypical platform, we use first-principles molecular dynamics with extended Hubbard functionals to follow, in real time and at finite temperature, the adiabatic evolution of Mn oxidation states and atomic sublattices. We find that polaron hopping and cation migration are dynamically decoupled, and \Mntwo/\Mnthree{} valence exchanges occur \revviii{frequently} while the intercalated ions remain confined to their sites: hopping is gated by the local oxygen framework.

\rev{In simulations, freezing} the six oxygen atoms of an individual \MnO{} octahedron selectively excludes that site from the hopping network, while increasing the fictitious oxygen mass systematically slows down electronic transport. Each valence exchange coincides with a concerted breathing of the donor and acceptor octahedra ($\Delta V \simeq 2$~\AA$^3$), with their volume asymmetry crossing zero at the transition, consistent with the geometric coincidence condition of the Holstein--Marcus picture. The temperature-dependent hopping rate yields an activation energy of $E_a = 0.20 \pm \red{0.03}$~eV, closely matching the elastic energy scale of the two coordination cages ($\lambda_V/4 \approx 0.20$~eV). These results identify the oxygen coordination cage, not the mobile cation sublattice, as the dynamical gate for small-polaron transport: a site can accept the electron only if its cage can breathe, the hopping rate is set by \revviii{temperature} and the mass of the oxygen framework, and the barrier is governed by cage elasticity. Selective constraints and mass scaling applied to first-principles dynamics provide a general route to \rev{establish links} between electronic and lattice degrees of freedom across redox-active materials, opening the door to \rev{framework engineering} in next-generation intercalation and mixed ionic--electronic conductors.
\end{abstract}

% =====================================================================
\section*{Introduction}

The concept of oxidation state (OS) is ubiquitous in physics and chemistry~\cite{Walsh2018}, and its evolution in time underlies the functioning of rechargeable batteries, in which redox reactions on the transition-metal (TM) ions of the cathode accompany the insertion and removal of lithium. Olivine phosphates Li$M$PO$_4$ ($M$\,=\,Fe, Mn) are prototypical intercalation cathodes~\cite{Padhi1997,Hautier2011}: their strong P--O covalent bonds confer thermal and chemical stability, and LiMnPO$_4$ is particularly attractive because of its higher redox potential ($\sim$4.1~V vs Li/Li$^+$) compared with LiFePO$_4$~\cite{Padhi1997,Martha2009}. Its practical use, however, has long been hampered by sluggish kinetics, which have been ascribed to an extremely low intrinsic electronic conductivity, lower than that of LiFePO$_4$ by several orders of magnitude~\cite{Delacourt2005,Martha2009}, combined with one-dimensional lithium diffusion along the $b$ channels~\cite{Morgan2004,Islam2005}.

In olivines, the extra electrons (or holes) introduced upon (de)lithiation do not delocalize in a band but self-trap on individual TM sites, together with a local distortion of the surrounding oxygen cage, forming small polarons~\cite{Holstein1959,Austin1969,Franchini2021,Zhou2004}. Electronic transport then occurs through thermally activated hops between neighbouring TM sites~\cite{Emin1969,Austin1969}. A central and still debated question is the extent to which these hops are coupled to the motion of the charge-compensating Li$^+$ ions. \revviii{Static first-principles calculations at zero temperature predict a sizeable electrostatic binding between polarons and Li ions or vacancies~\cite{Maxisch2006,Ong2011}, which raises the calculated hopping barrier of a bound polaron with respect to a free one~\cite{Ong2011};} on the other hand, M\"ossbauer spectroscopy on Li$_x$FePO$_4$ solid solutions has been interpreted in terms of coupled lithium-ion and electron mobility~\cite{Ellis2006}. In this coupled picture, the Li sublattice would effectively control electronic transport, and the polaron and the Li ion would migrate as a correlated pair.

Static calculations, such as nudged-elastic-band estimates of hopping barriers along predefined pathways~\cite{Maxisch2006,Ong2011,Johannes2012}, cannot by themselves resolve whether this binding governs finite-temperature transport. What is required is a description in which electrons, lithium ions and the oxygen framework evolve dynamically together at finite temperature, and in which the OS of every TM ion can be followed unambiguously in time. First-principles molecular dynamics (FPMD)~\cite{Car1985} offers such a description, provided the electronic-structure method localizes the excess charges correctly. Standard (semi)local functionals fail in this respect because of self-interaction errors, delocalizing the extra electrons over all TM sites and producing ``unchemical'' fractional OSs~\cite{Timrov2022,Malica2025}. Hybrid functionals~\cite{Heyd2003} correct this\rev{, but are typically unaffordable for FPMD simulations long enough to yield meaningful statistics.} Extended Hubbard functionals (\UV)~\cite{Campo2010}, with on-site $U$ and inter-site $V$ parameters computed from first principles~\cite{Timrov2018,Timrov2021,Timrov2022hp}, provide accurate electronic structures and intercalation voltages for olivines~\cite{Cococcioni2019,Timrov2022}, and \UV-FPMD captures sharp, ``digital'' changes of the Mn OS during the \revviii{adiabatic} dynamics of \LMPO{}~\cite{Malica2025}.

Here, we exploit this capability to \rev{analyze} the microscopic mechanism of polaron hopping in \LMPO{}. By combining unconstrained and selectively constrained FPMD, fictitious isotope substitution, \rev{analysis} of the electron--lattice coupling and the temperature dependence of the hopping rate, we show that (i) polaron hopping is dynamically decoupled from lithium motion\revLi{, proceeding on timescales much shorter than those of lithium migration}, (ii) it is gated by the local oxygen cage of each Mn site, (iii) the donor--acceptor volume asymmetry crosses zero at the instant of transfer through a geometrically symmetric coincidence transition state, and (iv) the resulting activation energy ($E_a = 0.20\pm\red{0.03}$~eV) is quantitatively consistent with the elastic reorganization energy of two breathing cages ($\sim$0.20~eV) and matches the barrier of a free electron polaron. The oxygen framework thus emerges as the controlling degree of freedom for electronic transport in these polyanionic cathodes.

% =====================================================================
\section*{Results}

\subsection*{Following oxidation states in time}

In \LMPO{} (orthorhombic, space group $Pnma$), each Mn ion sits at the centre of a distorted \MnO{} octahedron; the octahedra share corners in the $bc$ plane and are linked by PO$_4$ tetrahedra, while Li$^+$ ions occupy one-dimensional channels along $b$~\cite{Padhi1997,Morgan2004}. Each Li inserted in MnPO$_4$ brings one Li$^+$ ion and one electron, which reduces one Mn$^{3+}$ ion to Mn$^{2+}$. Following the approach of \rev{Ref.}~\citenum{Sit2011}, we determine the OS of each Mn from the eigenvalues of its 3$d$ occupation matrix, and use as a compact descriptor the L\"owdin occupation $n$, i.e.\ the sum of the 3$d$ occupation eigenvalues in both spin channels~\cite{Lowdin1950,Malica2025}. We emphasize that $n$ is a hybridized L\"owdin occupation: the contrast of $\Delta n \approx 0.20$~$e$ between \Mnthree{} ($n\simeq 5.00$) and \Mntwo{} ($n\simeq 5.20$) in \UV{}~\cite{Malica2025} reflects covalent hybridization with coordinating oxygen 2$p$ states, while the integer change in oxidation state is unambiguously defined by the occupation-matrix eigenvalues~\cite{Sit2011,Malica2025}. \rev{Thermal motion makes $n$ fluctuate} around these two values; we therefore define two \rev{intervals}, $n<5.05$ (Mn$^{3+}$) and $n>5.15$ (Mn$^{2+}$), separated by a hysteresis band that suppresses spurious recrossings, and count an OS change (``hop'') only when the occupation enters the opposite \rev{interval} and remains there for at least a dwell time $\tdw$ (Methods). \rev{We underscore} that the transitions observed in \UV-FPMD are \emph{adiabatic} changes of the electronic ground state driven by the evolving atomic configuration~\cite{Malica2025}; we do not attempt to describe non-adiabatic tunnelling~\cite{Marcus1985,Sit2006}, which would take place on the same configurations but with an electronically reduced transmission probability.

% ---------------------------------------------------------------------
\subsection*{Polaron hopping proceeds while lithium stays put}

\begin{figure}[t]
  \centering
  \includegraphics[width=\textwidth]{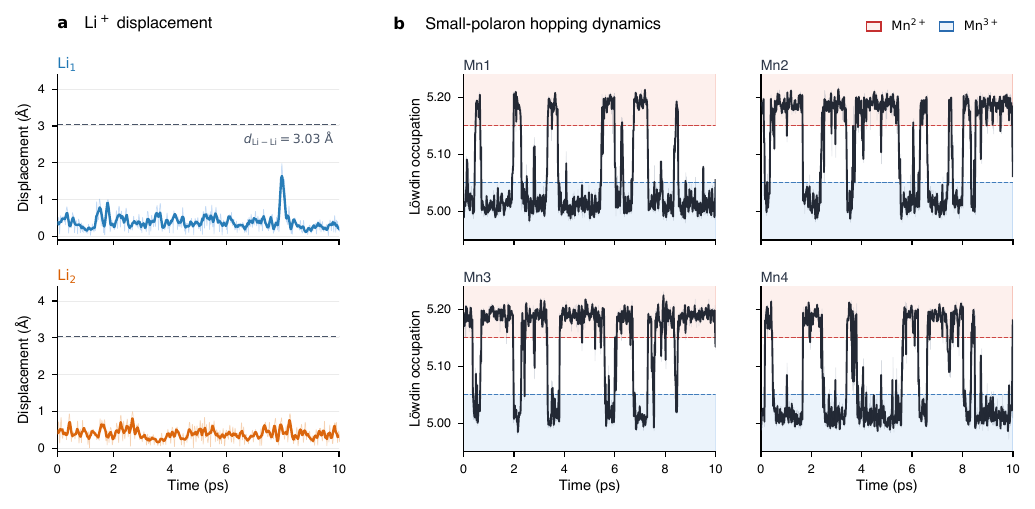}
  \caption{\textbf{Decoupling of lithium motion and small-polaron dynamics in Li$_{0.5}$MnPO$_4$.}
  \textbf{a}, Displacements $|\mathbf{R}(t)-\mathbf{R}(0)|$ of the two Li$^+$ ions (Li$_1$, Li$_2$) during a 10~ps \UV-FPMD trajectory at $T=900$~K (unit cell of \revviii{26 atoms, with two Li$^+$ ions and two extra electrons)}. Faint lines: raw data; solid lines: 0.1 ps running average. The dashed line marks the \red{distance between nearest-neighbour Li sites, $d_{\mathrm{Li-Li}}=3.03$~\AA}. Both ions only vibrate within their sites.
  \textbf{b}, L\"owdin 3$d$ occupations of the four Mn sites over the same time window ($\delta t=4$~fs). Shaded bands indicate the Mn$^{2+}$ ($n>5.15$) and Mn$^{3+}$ ($n<5.05$) basins, dashed lines the corresponding thresholds. Faint lines: raw data; solid lines: median filter over a 20 fs (5-frame) window. The Mn oxidation states change continuously through polaron hops while the Li sublattice is static.}
  \label{fig:decoupling}
\end{figure}

Figure~\ref{fig:decoupling} shows a 10~ps \UV-FPMD trajectory of Li$_{0.5}$MnPO$_4$ at $T=900$~K, in a unit cell \revviii{of 26 atoms} containing two Li$^+$ ions and two extra electrons. The two Li ions (Fig.~\ref{fig:decoupling}a) never leave their sites: their displacements $|\Delta\mathbf{R}(t)|=|\mathbf{R}(t)-\mathbf{R}(0)|$ remain below 1~\AA{} (with a single transient excursion below 2~\AA), far from the \red{distance between nearest-neighbour Li sites, $d_{\mathrm{Li-Li}}=3.03$~\AA}. Over the same time window, the L\"owdin occupations of the four Mn sites (Fig.~\ref{fig:decoupling}b) switch repeatedly and sharply between the Mn$^{3+}$ and Mn$^{2+}$ basins, with electrons always conserved (two Mn$^{2+}$ and two Mn$^{3+}$ at any time outside the short transition windows). Polaron hopping is therefore active on the picosecond timescale even when the Li sublattice is completely static \rev{(i.e., oscillating at equilibrium and not hopping)}. \revLi{The high temperature is chosen to obtain sufficient hopping statistics within a single trajectory. On the full 38~ps of this run the two Li ions perform only three site changes, and none occurs in the 600 and 700~K trajectories used below (32.7 and 41.7~ps), whereas the polaron keeps hopping at all temperatures.} \revviii{Hence, the hopping does not require Li migration.}

\begin{figure}[t]
  \centering
  \includegraphics[width=\textwidth]{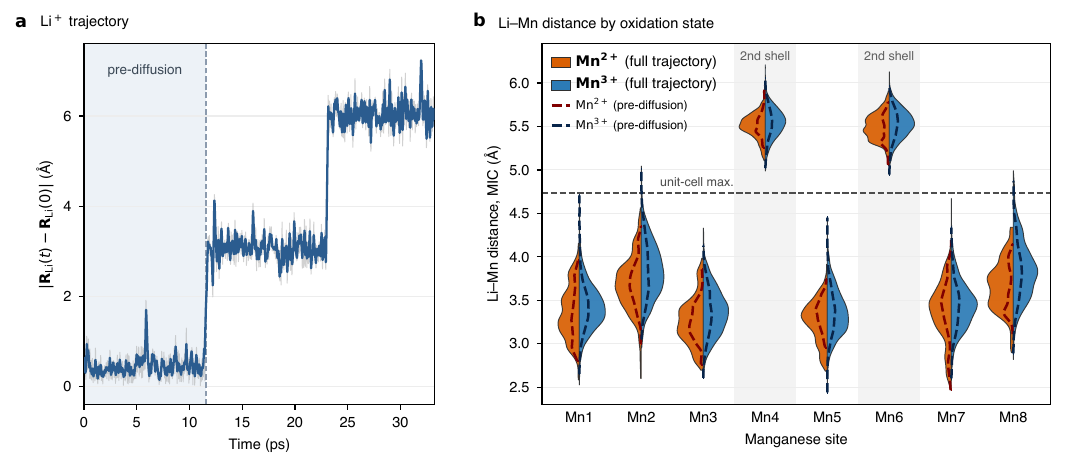}
  \caption{\revviii{\textbf{Li migration and polaron hopping in the Li$_{1/8}$MnPO$_4$ supercell at 900~K.}
  $1\times1\times2$ supercell (49 atoms: 8 Mn, 8 P, 32 O, 1 Li; $a\times b\times c\approx9.94\times6.03\times9.63$~\AA) with one Li$^+$ ion and one polaron, 33.3~ps (8,314 frames, $\delta t=4$~fs).
  \textbf{a}, Displacement of the Li$^+$ ion. The plateaus at $\approx$0, 3.1 and 6.0~\AA{} \red{correspond to two successive jumps between nearest-neighbour Li sites ($\approx$3~\AA{} each)}. The shaded window ($t\leq11.5$~ps) marks the pre-diffusion stage, before the first Li jump. Faint line: raw data; solid line: 100~fs running average.
  \textbf{b}, Split-violin distributions of the instantaneous Li--Mn distance under the minimum-image convention (MIC, range 2.41--6.21~\AA) for each Mn site, conditioned on the oxidation state of that site: Mn$^{2+}$ (orange, $n>5.137$) and Mn$^{3+}$ (blue, $n<5.035$), with thresholds set by the same convention as in the unit cell (Methods). Filled violins: full trajectory; dashed lines: contribution of the pre-diffusion stage (dark red, Mn$^{2+}$; dark blue, Mn$^{3+}$), scaled by its share of the frames. Grey bands: second-shell sites (Mn4, Mn6). Horizontal dashed line: largest Li--Mn MIC distance in the unit cell (4.73~\AA; Supplementary Fig.~\ref{fig:SIunitcell}).}}
  \label{fig:association}
\end{figure}

\revviii{To test whether the polaron is pinned to the Li$^+$ ion, we analysed two trajectories with a single Li$^+$ ion and a single polaron at 900~K: the Li$_{0.25}$MnPO$_4$ unit cell of 25 atoms (27.8~ps) and a $1\times1\times2$ Li$_{1/8}$MnPO$_4$ supercell of 49 atoms (33.3~ps). For every Mn site we evaluated the distribution of instantaneous Li--Mn distances under the minimum-image convention (MIC, i.e., taking the shortest distance across periodic--boundary conditions), separately for the frames in which that site is Mn$^{2+}$ and for those in which it is Mn$^{3+}$. In the unit cell (Supplementary Fig.~\ref{fig:SIunitcell}), the Li$^+$ ion \red{jumps six times between nearest-neighbour Li sites}. Every Mn site lies within the first coordination shell of the Li (2.35--4.73~\AA), and both oxidation states occupy the full distance range at every site. The supercell (Fig.~\ref{fig:association}) adds two Mn sites (Mn4, Mn6) at $\approx$5.5~\AA{} from the Li, the second-shell distance at which static calculations place the largest part of the Li$^+$--polaron binding~\cite{Maxisch2006}. The polaron also visits these sites, and their Mn$^{2+}$ and Mn$^{3+}$ distributions overlap. This already holds in the first 11.5~ps, before the Li$^+$ ion leaves its initial site (shaded window in Fig.~\ref{fig:association}a; dashed profiles in Fig.~\ref{fig:association}b): during this stage the polaron visits all eight Mn sites, including Mn4 and Mn6.}

% ---------------------------------------------------------------------
\subsection*{The local oxygen cage gates the hopping}

The results above show that Li motion is not required for polaron hopping; we now ask which degrees of freedom \emph{are} required. To this end we performed two FPMD simulations, at $T=900$~K and for 13.46~ps each, of a $1\times1\times2$ supercell (49 atoms: eight Mn, eight P, 32 O and one Li, i.e.\ Li$_{1/8}$MnPO$_4$ with a single polaron). In the first, all atoms are free; in the second, the six oxygen atoms forming the octahedron of one Mn site (Mn2) are frozen at their initial positions, while all other atoms, including Mn2 itself, evolve freely (Methods).

\begin{figure}[p]
  \centering
  \includegraphics[width=\textwidth]{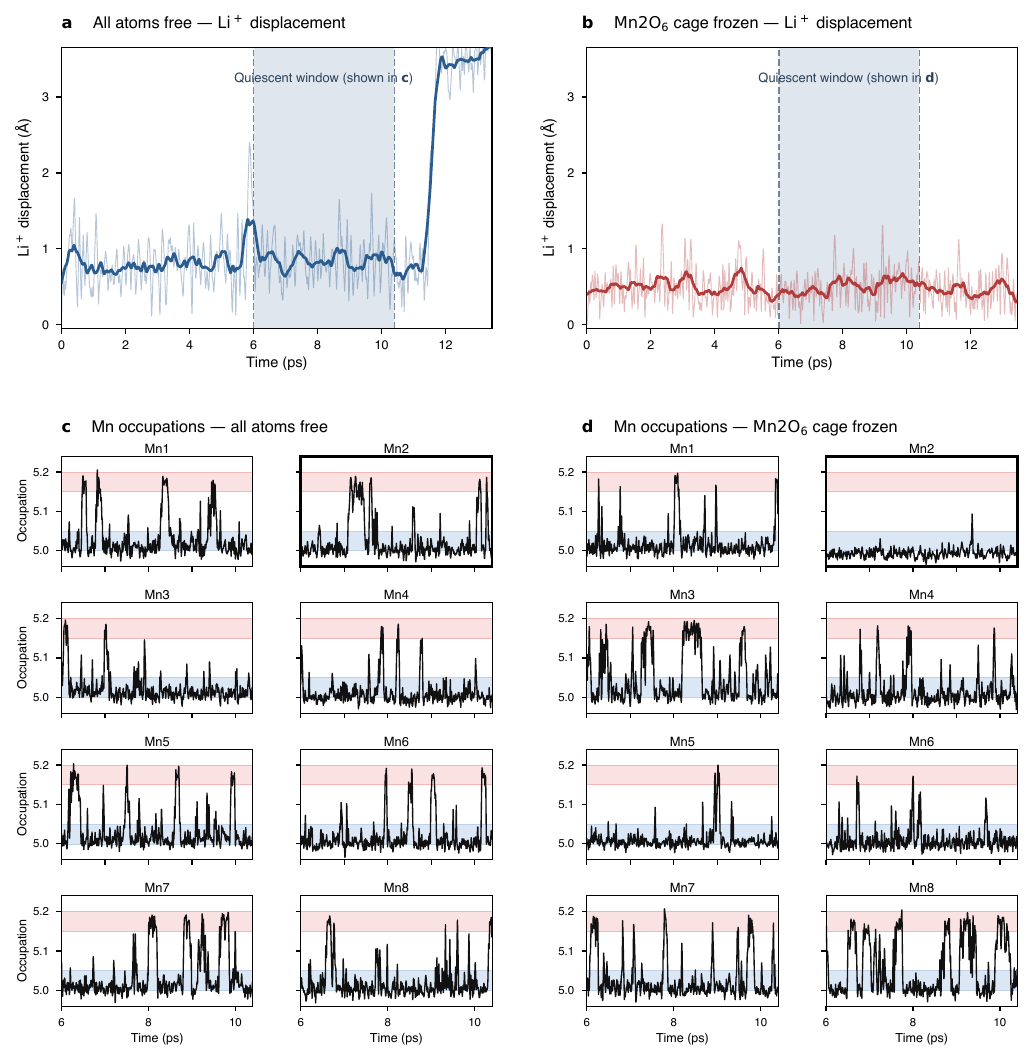}
  \caption{\textbf{Site-selective freezing of the oxygen cage.}
  \UV-FPMD of Li$_{1/8}$MnPO$_4$ ($1\times1\times2$ supercell, 49 atoms\revviii{: 8 Mn, 8 P, 32 O, 1 Li; $a\times b\times c\approx9.94\times6.03\times9.63$~\AA}, one Li$^+$ ion and one polaron) at $T=900$~K, 13.46~ps, $\delta t=4$~fs.
  \textbf{a}, Li$^+$ displacement with all atoms free (faint: raw data; solid: 0.4~ps running average). The Li ion vibrates within its site until $t\approx11.4$~ps, when it jumps to a neighbouring site. The grey band (6.0--10.4~ps) marks the window shown in \textbf{c} and \textbf{d}.
  \textbf{b}, Same as \textbf{a} with the six oxygen atoms of the Mn2O$_6$ octahedron frozen at their initial positions.
  \textbf{c}, L\"owdin occupations of the eight Mn sites with all atoms free, in the window 6.0--10.4~ps. Blue and red bands indicate the Mn$^{3+}$ and Mn$^{2+}$ basins. The polaron hops among all sites, including Mn2 (framed).
  \textbf{d}, Same as \textbf{c} with the Mn2O$_6$ cage frozen: Mn2 (framed) remains Mn$^{3+}$ for the entire trajectory, while hopping continues among the other seven sites at an unchanged overall rate.}
  \label{fig:constraints}
\end{figure}

In the unconstrained run (Fig.~\ref{fig:constraints}a,c), the Li$^+$ ion vibrates within its site for $\sim$11.4~ps before performing a single jump; during the quiescent window highlighted in Fig.~\ref{fig:constraints} (6.0--10.4~ps), the polaron hops continuously among all eight Mn sites, including Mn2. Over the whole trajectory we count \red{154} OS transitions ($\tdw\geq12$~fs), i.e.\ \red{77} polaron hops, with Mn2 participating in \red{25} of them. When the Mn2O$_6$ cage is frozen (Fig.~\ref{fig:constraints}b,d), Mn2 remains Mn$^{3+}$ throughout the entire trajectory, with zero transitions and strongly reduced occupation fluctuations, whereas the polaron keeps hopping among the other seven sites; the total number of transitions (\red{146}) is essentially unchanged. Freezing a single coordination cage is therefore sufficient to remove the corresponding site from the hopping network, without perturbing hopping elsewhere. 

Because the frozen cage retains the geometry of the initial Mn$^{3+}$ configuration ($V\approx11.2$~\AA$^3$), this result demonstrates that a site can host the polaron only if its oxygen cage can expand to accommodate the larger Mn$^{2+}$ ion ($V\approx13.0$~\AA$^3$, see below). Freezing a cage in the expanded Mn$^{2+}$ geometry would be the complementary test; we expect it to stabilize the polaron on that site and suppress hopping away from it, since the coincidence condition could then be met only through large expansions of a neighbouring cage. We note that the Li ion did not jump in the constrained run; given that only a single jump was observed in the unconstrained one, we refrain from drawing conclusions on Li mobility from this comparison.

A \rev{complementary probe} of the role of the oxygen framework is provided by fictitious isotope substitution. In classical molecular dynamics on an electronic Born--Oppenheimer potential energy surface, altering the atomic masses leaves the potential energy landscape and all static activation barriers strictly invariant; any variation in the hopping rate thus isolates the purely dynamical effect of framework inertia. We note that this classical dynamical effect is physically distinct from the quantum polaronic isotope effect observed in perovskite manganites~\cite{Zhao1996}, where oxygen substitution ($^{16}\mathrm{O}\to{}^{18}\mathrm{O}$) renormalizes the electronic bandwidth and polaronic binding through zero-point lattice fluctuations. Here we isolate framework inertia by setting the fictitious oxygen mass $m_{\mathrm{O}}$ to 19 (F), 32 (S) and 79~u (Se).

\begin{figure}[t]
  \centering
  \includegraphics[width=\textwidth]{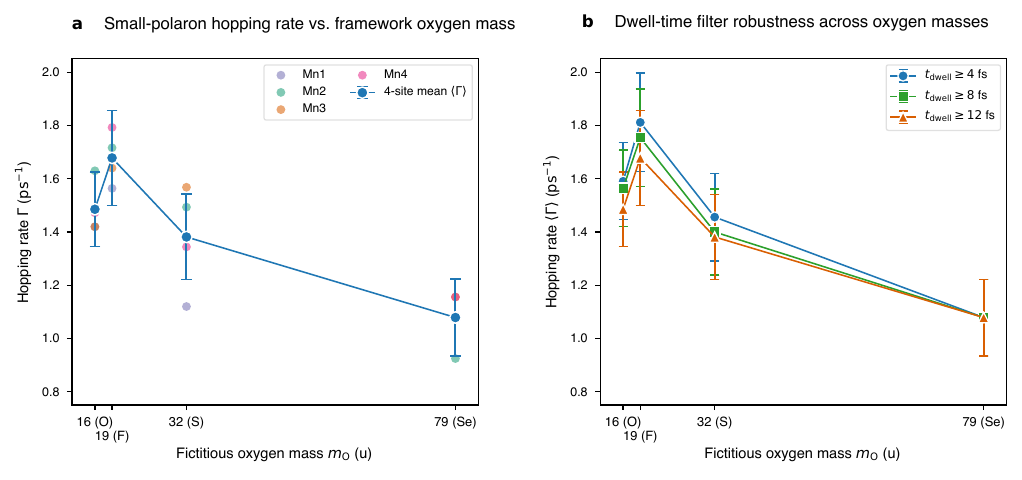}
  \caption{\textbf{Fictitious oxygen-isotope effect on polaron hopping.}
  \UV-FPMD of Li$_{0.5}$MnPO$_4$ \revviii{(unit cell, 26 atoms: 4 Mn, 4 P, 16 O, 2 Li; $a\times b\times c\approx10.15\times6.07\times4.81$~\AA)} at $T=900$~K with oxygen mass $m_{\mathrm{O}}=16$ (O), 19 (F), 32 (S) and 79~u (Se), on an unchanged Born--Oppenheimer potential-energy surface (trajectories of \red{26.0}--38.0~ps, $\delta t=4$~fs).
  \textbf{a}, Polaron hopping rate as a function of $m_{\mathrm{O}}$. Large symbols: four-site average $\langle\Gamma\rangle=\frac{1}{4}\sum_i N_i/t_{\mathrm{tot}}$ ($\tdw\geq12$~fs), with Poisson uncertainties \red{$\langle\Gamma\rangle/\sqrt{N_{\mathrm{hop}}}$, where $N_{\mathrm{hop}}=N_{\mathrm{tot}}/2$ is the number of hops}; small symbols: individual Mn sites.
  \textbf{b}, Hopping rate $\langle\Gamma\rangle$ evaluated for dwell-time thresholds $\tdw\geq4$, 8 and 12~fs.}
  \label{fig:isotope}
\end{figure}

Figure~\ref{fig:isotope}a shows the polaron hopping rate $\langle\Gamma\rangle$ for Li$_{0.5}$MnPO$_4$ at 900~K (trajectories of \red{26.0}--38.0~ps). \red{The rate at $m_{\mathrm{O}}=19$~u ($1.68\pm0.18$~ps$^{-1}$) is comparable within statistical uncertainty to that at 16~u ($1.49\pm0.14$~ps$^{-1}$), whereas heavier oxygen masses lower it, down to $1.08\pm0.14$~ps$^{-1}$ at 79~u, a reduction of $\sim$25--35\%. At 79~u all four Mn sites hop more slowly than at 16 and 19~u, and the decrease is found for all dwell-time criteria (Fig.~\ref{fig:isotope}b), excluding artefacts of the counting procedure. Figure~\ref{fig:isotope} thus indicates that polaron hopping kinetics is slowed down by the inertia of the oxygen framework.}

% ---------------------------------------------------------------------
\subsection*{Concerted electron--lattice dynamics and Marcus coincidence}

\begin{figure}[p]
  \centering
  \includegraphics[width=0.93\textwidth]{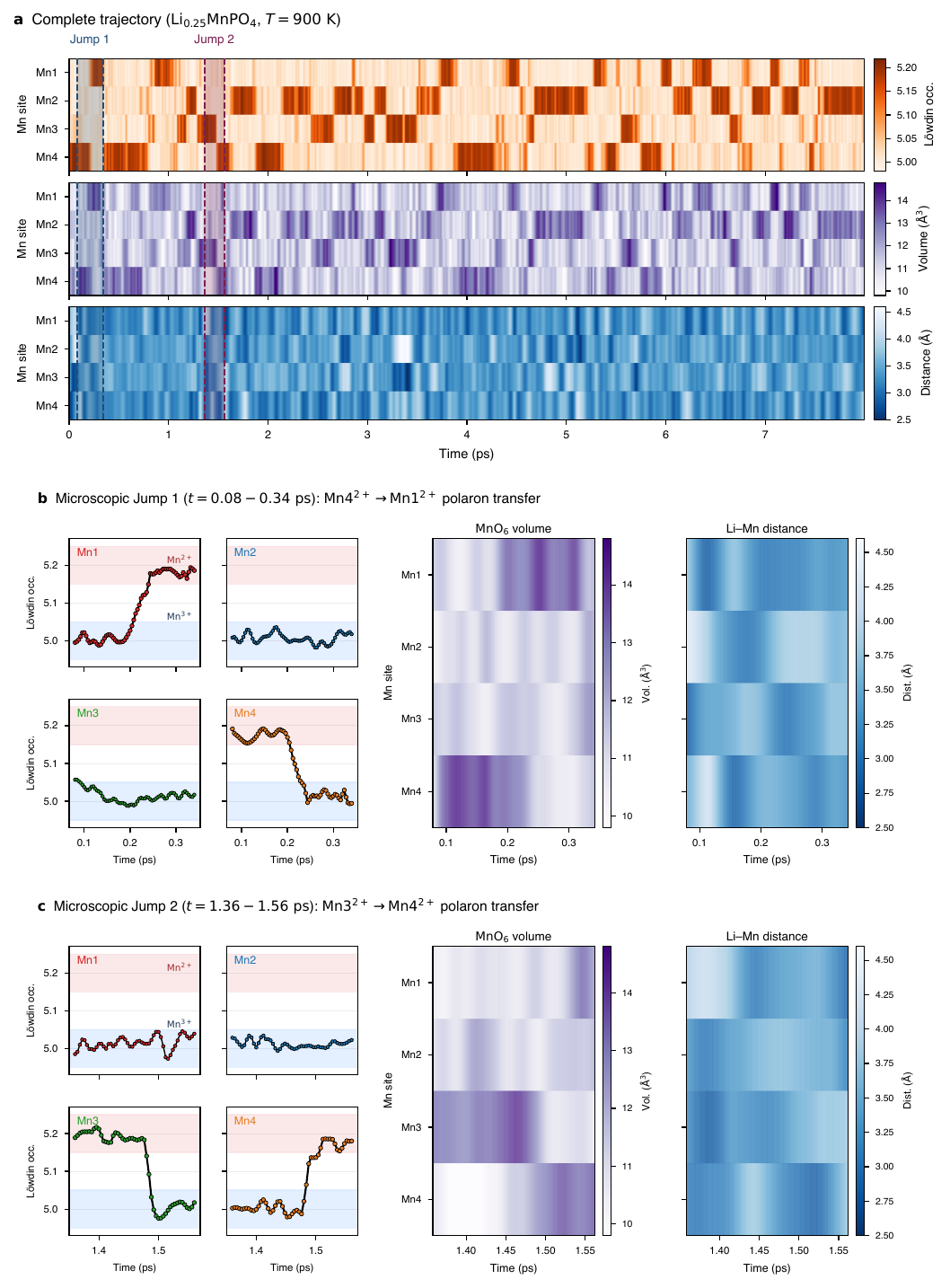}
  \caption{\textbf{Structural fingerprint of polaron hops in Li$_{0.25}$MnPO$_4$ at $T=900$~K.} \revviii{Unit cell, 25 atoms: 4 Mn, 4 P, 16 O, 1 Li; $a\times b\times c\approx9.93\times6.04\times4.84$~\AA, one Li$^+$ ion and one polaron.}
  \textbf{a}, Heatmaps over 8~ps of the Mn L\"owdin occupations (top), \MnO{} octahedral volumes from convex-hull tessellation (middle) and Li--Mn distances (minimum-image convention; bottom). Shaded windows mark the two events shown in \textbf{b} and \textbf{c}.
  \textbf{b}, Zoom on jump~1 ($t=0.08$--0.34~ps): L\"owdin occupations of the four Mn sites (left), with the polaron transferred from Mn4 to Mn1, octahedral volumes (centre) and Li--Mn distances (right).
  \textbf{c}, Same for jump~2 ($t=1.36$--1.56~ps), with the polaron transferred from Mn3 to Mn4. In both events the donor cage contracts and the acceptor cage expands by $\sim$2~\AA$^3$ in step with the electronic transition, while the Li--Mn distances do not change appreciably.}
  \label{fig:fingerprint}
\end{figure}

Figure~\ref{fig:fingerprint} resolves individual hopping events in Li$_{0.25}$MnPO$_4$ at 900~K. The overview heatmaps (Fig.~\ref{fig:fingerprint}a) show that the Mn OS pattern (top) is mirrored by the \MnO{} octahedral volumes (middle), whereas no corresponding pattern is visible in the Li--Mn distances (bottom). The zooms on two representative events (Fig.~\ref{fig:fingerprint}b,c) display the anatomy of a hop: within a few tens of femtoseconds, the occupation of the donor site drops from $\sim$5.18 to $\sim$5.00 while that of the acceptor rises from $\sim$5.00 to $\sim$5.18; at the same time, the donor octahedron contracts and the acceptor octahedron expands by $\sim$2~\AA$^3$, while the Li--Mn distances remain essentially unaffected. \rev{Adiabatic electron} transfer and cage breathing thus occur as a single concerted event, involving the donor--acceptor pair of octahedra\rev{; of course, the matrix element between diabatic states~\cite{Sit2006,Migliore2009} would ultimately renormalize the reaction rate and forbid long-distance tunnelling}.

\begin{figure}[!htbp]
  \centering
  \includegraphics[width=\textwidth]{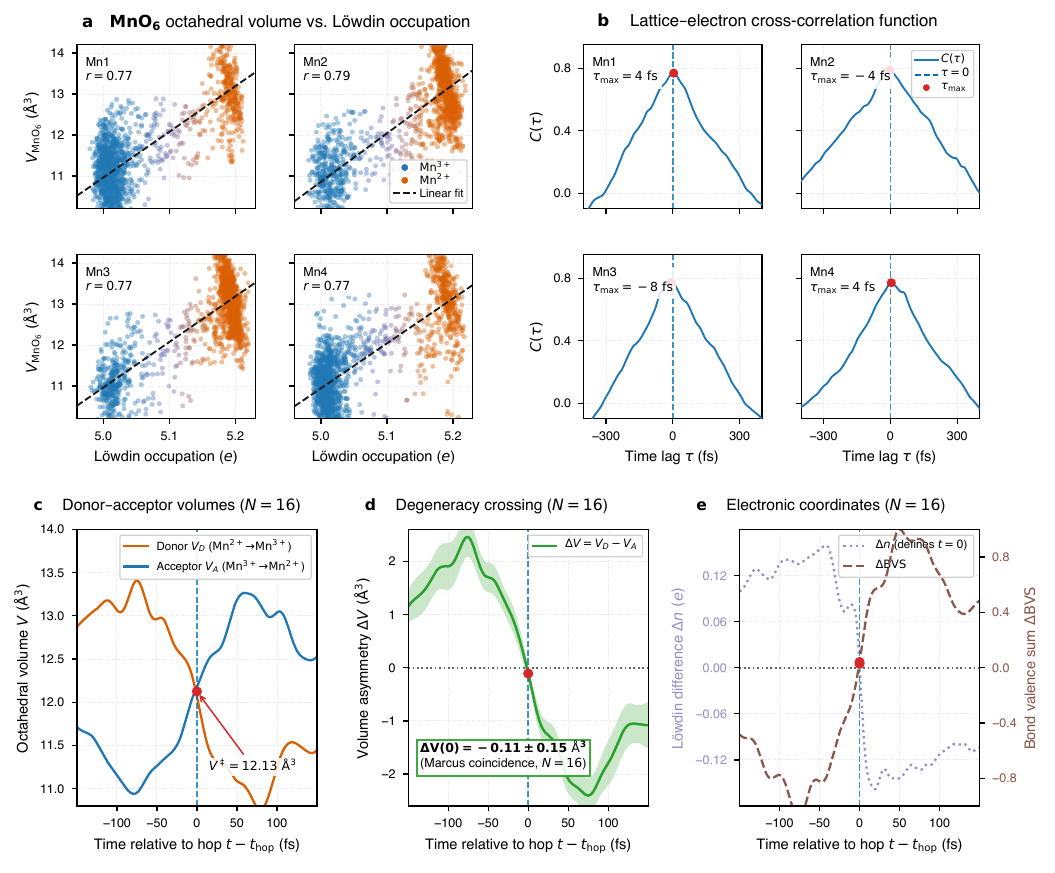}
  \caption{\textbf{Concerted electron--lattice dynamics and Holstein--Marcus coincidence in Li$_{0.5}$MnPO$_4$ at $T=900$~K.} \revviii{Unit cell, 26 atoms: 4 Mn, 4 P, 16 O, 2 Li; $a\times b\times c\approx10.15\times6.07\times4.81$~\AA, two Li$^+$ ions and two polarons.}
  \textbf{a}, \MnO{} volume versus L\"owdin occupation for the four Mn sites (6~ps, 1,500 frames), showing bimodal clustering for Mn$^{3+}$ (blue) and Mn$^{2+}$ (orange) and linear regression fits (dashed black lines, Pearson $r=0.77$--0.79).
  \textbf{b}, Normalized cross-correlation $C(\tau)$ between volume and occupation; the maximum (red dot) \red{lies within $\pm8$~fs of $\tau=0$ for all sites, with $C(0)$ within 0.002 of the peak value,} confirming that octahedral breathing tracks the electronic state without \red{appreciable} phase lag.
  \textbf{c}, Hop-aligned donor ($V_D$) and acceptor ($V_A$) octahedral volumes over $[-150, +150]$~fs ($\delta t = 0.5$~fs resolution, $N = 16$ hops), crossing at the transition-state volume $V^\ddagger \approx 12.12$~\AA$^3$ at $t = 0$.
  \textbf{d}, \red{Event-averaged cage volume asymmetry $\Delta V(t) = V_D(t) - V_A(t)$ approaching zero near the electronic switch ($\Delta V(0) = -0.11 \pm 0.15$~\AA$^3$, mean $\pm$ standard error over $N = 16$ hops; shaded band: $\pm 1\sigma$, $\sigma = 0.61$~\AA$^3$), consistent with the Holstein--Marcus coincidence condition.}
  \textbf{e}, Simultaneous electronic transfer: L\"owdin occupation difference $\Delta n(t) = n_D - n_A$ (dotted line; $\Delta n = 0$ defines $t = 0$) and bond-valence sum mismatch $\Delta\mathrm{BVS}(t) = \mathrm{BVS}_D - \mathrm{BVS}_A$ (dashed line) crossing zero at $t = 0$.}
  \label{fig:correlation}
\end{figure}

This concerted behaviour is quantified in Fig.~\ref{fig:correlation} for Li$_{0.5}$MnPO$_4$ at 900~K (6~ps, 1,500 frames). The instantaneous octahedral volume and the L\"owdin occupation are strongly correlated on every site (Pearson $r=0.77$--0.79; Fig.~\ref{fig:correlation}a), with average volumes of $\sim$11.2~\AA$^3$ for Mn$^{3+}$ and $\sim$13.0~\AA$^3$ for Mn$^{2+}$; the larger, Jahn--Teller-inactive high-spin Mn$^{2+}$ ion requires a larger and more regular cage than the Jahn--Teller-active Mn$^{3+}$~\cite{Reed2004}. The normalized cross-correlation $C(\tau)$ between the two time series (Fig.~\ref{fig:correlation}b) peaks \red{within $\pm8$~fs (two time steps) of zero lag for all sites, with $C(0)$ within 0.002 of the maximum; on the timescale of cage reorganization (Fig.~\ref{fig:correlation}e) the lag between the two series is therefore zero within resolution}. 

In Born--Oppenheimer dynamics, the electronic ground state is an instantaneous functional of the nuclear coordinates. Therefore, a vanishing phase lag between $n(t)$ and $V(t)$ is physically expected for any structural descriptor that captures the reaction coordinate; its significance is that the \MnO{} octahedral volume, and even more closely the bond-valence sum (Supplementary Figs.~\revviii{S2--S4}), directly tracks the electronic coordinate rather than a delayed structural relaxation \rev{(i.e., when the oxygens breathe, the electron jumps)}. To test the microscopic transfer mechanism, we \rev{aligned all transitions at the instant of electron transfer and averaged them over a trajectory with a shorter time step} ($\delta t = 0.5$~fs, 10,216 frames, 5.11~ps; Fig.~\ref{fig:correlation}c--e). Aligning the trajectories at the transfer time ($t = 0$), the donor and acceptor cage volumes $V_D(t)$ and $V_A(t)$ evolve symmetrically towards each other, crossing at a transition-state volume $V^\ddagger \approx 12.12$~\AA$^3$ midway between the \Mnthree{} and \Mntwo{} cages (Fig.~\ref{fig:correlation}c). The donor--acceptor volume asymmetry coordinate:
\begin{equation}
  \Delta V(t) = V_D(t) - V_A(t)
\end{equation}
\red{approaches zero near the electronic switch ($\Delta V(0) = -0.11 \pm 0.15$~\AA$^3$, averaged over $N = 16$ hops; Fig.~\ref{fig:correlation}d). In the event average, the donor and acceptor cage volumes become nearly equal near the electronic switch, consistent with a predominantly breathing reaction coordinate and the Holstein--Marcus coincidence condition~\cite{Marcus1956,Holstein1959}, with the two cages reorganizing concertedly.} While the electronic occupation crossing ($\Delta n(0) = 0$) defines the temporal origin $t = 0$, the crucial physical finding is the simultaneous vanishing of the purely geometric asymmetries $\Delta V(0) = -0.11 \pm 0.15$~\AA$^3$ and $\Delta\mathrm{BVS}(0) \approx 0$, accompanied by the strictly symmetric, $\sim\pm50$~fs concerted reorganization of both cages around the transfer (Fig.~\ref{fig:correlation}e). \rev{This shows that the electron transfers through a geometrically symmetric configuration, consistent with the coincidence state required for adiabatic polaronic transport.} \rev{In the language of Marcus theory, the hop takes place where the diabatic free-energy curves of the electron localized on donor and acceptor cross; the electronic coupling between the two sites splits these curves, and in the adiabatic limit followed here the electron transfers with unit probability whenever the crossing region is reached.}

% ---------------------------------------------------------------------
\subsection*{\red{Activation energy of local polaron hopping}}

\begin{figure}[!htbp]
  \centering
  \includegraphics[width=\textwidth]{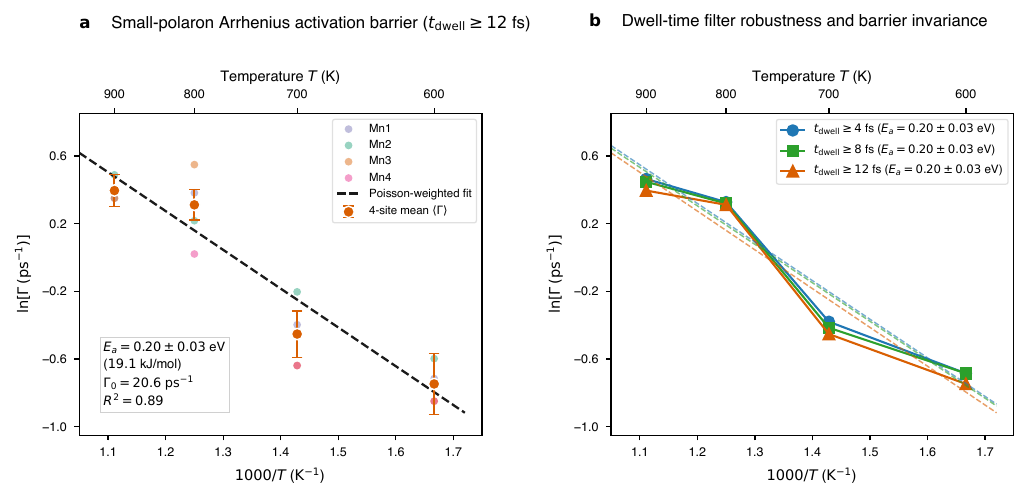}
  \caption{\textbf{Thermally activated polaron hopping in Li$_{0.5}$MnPO$_4$.} \revviii{Unit cell, 26 atoms: 4 Mn, 4 P, 16 O, 2 Li; $a\times b\times c\approx10.15\times6.07\times4.81$~\AA, two Li$^+$ ions and two polarons.}
  \textbf{a}, Arrhenius plot of the hopping rate from FPMD at 600--900~K (trajectories of 32.7--45.1~ps\rev{, $\tdw\geq12$~fs}). Small symbols: individual Mn sites; large symbols: four-site average $\langle\Gamma\rangle$, equal to the hopping rate per polaron, with Poisson uncertainties. The dashed line is a \rev{Poisson-weighted} linear fit of $\ln\langle\Gamma\rangle$ versus $1/T$, yielding $E_a=0.20\pm\red{0.03}$~eV and \rev{$\Gamma_0 \approx 2\times10^{13}$~s$^{-1}$ ($R^2=0.89$)}.
  \textbf{b}, Arrhenius plots for dwell-time thresholds $\tdw\geq4$, 8 and 12~fs, \rev{showing a $\sim$6\% rate variation between 4 and 12~fs and yielding a consistent activation energy $E_a=0.20\pm\red{0.03}$~eV for all three thresholds}.}
  \label{fig:arrhenius}
\end{figure}

Finally, we determined the temperature dependence of the hopping rate from FPMD trajectories of Li$_{0.5}$MnPO$_4$ at 600, 700, 800 and 900~K, of 32.7--45.1~ps each (\rev{157.5~ps in total}). Since every hop changes the OS of exactly two Mn sites, and the cell hosts two polarons, the average of the site transition rates, $\langle\Gamma\rangle=\frac{1}{4}\sum_i N_i/t_{\mathrm{tot}}$, coincides with the hopping rate per polaron (Methods). \rev{The Arrhenius plot (Fig.~\ref{fig:arrhenius}a, $\tdw \geq 12$~fs) yields an activation energy $E_a=0.20\pm\red{0.03}$~eV and a prefactor $\Gamma_0\simeq2\times10^{13}$~s$^{-1}$ ($R^2=0.89$ from a Poisson-weighted fit).} The prefactor, which includes the number of accessible neighbouring sites and the effective attempt frequency, is of the order of $10^{13}$~s$^{-1}$, as expected for phonon-assisted adiabatic hopping~\cite{Emin1969,Austin1969}. Crucially, changing the dwell-time criterion from 4 to 12~fs lowers the absolute rates \rev{by only $\sim$6\%}, as transient excursions are discarded, \rev{and leaves the slope unchanged ($E_a=0.20\pm\red{0.03}$~eV across $\tdw\geq4$, 8 and 12~fs; Fig.~\ref{fig:arrhenius}b)}: the activation energy is an intrinsic property of the dynamics rather than of the counting procedure.

This barrier can be estimated, without adjustable parameters, from the equilibrium fluctuations of individual cages. Treating each cage volume as a harmonic coordinate, equipartition gives the stiffness:
\begin{equation}
  K = \frac{k_B T}{\langle \delta V^2 \rangle} = 0.182\text{ eV/\AA}^6,
\end{equation}
with $\langle \delta V^2 \rangle = 0.427$~\AA$^6$ (0.373 and 0.480~\AA$^6$ for \Mnthree{} and \Mntwo{} cages, respectively; neither environment is rigid, and their fluctuations have comparable amplitude). A hop requires the donor cage to contract and the acceptor cage to expand by $\Delta V = 2.115$~\AA$^3$, each at an elastic cost of $\frac{1}{2}K(\Delta V)^2$, so that the total reorganization energy of the two participating cages is:
\begin{equation}
  \label{eq:lambda_V}
  \lambda_V = 2 \times \frac{1}{2}K(\Delta V)^2 = K(\Delta V)^2 = 0.81\text{ eV},
\end{equation}
and the symmetric Marcus barrier is:
\begin{equation}
  E_a = \frac{\lambda_V}{4} \approx 0.20\text{ eV}.
\end{equation}
\red{This independent-cage harmonic estimate of the reorganization energy~\cite{Marcus1956,Holstein1959} is comparable to the Arrhenius activation energy of $0.20\pm0.03$~eV obtained under an independent-hop analysis (a quantitative Marcus description additionally requires diabatic energy gaps, covariance and electronic-coupling estimates).} Nonetheless, the conclusion is that the elasticity of the \MnO{} cages alone sets a barrier of the observed magnitude. \revLi{The cage stiffness is, moreover, temperature independent: \revviii{in our simulations,} between 600 and 900~K, $\langle\delta V^2\rangle$ grows linearly with $T$, giving $K = 0.17$--$0.18$~eV/\AA$^6$ and $\lambda_V/4 = 0.19$--$0.20$~eV at each temperature.}

\red{The observed $0.20$~eV barrier} is quantitatively consistent with the migration barrier computed with hybrid functionals (HSE06) for a free electron polaron in MnPO$_4$ (196~meV)~\cite{Ong2011}\rev{, a functional whose accuracy for olivines is only slightly worse than that of \UV{}~\cite{Cococcioni2019,Timrov2022}.} \revLi{Because every Mn site in this composition has two Li ions within 4.5~\AA, the polaron never has to escape from a bound site: the measured barrier characterizes hopping between Li-adjacent sites, not the escape of a trapped polaron.}

Analysis of the hopping pairs reveals that in collinear antiferromagnetic FPMD, hops occur both within the same spin sublattice and between opposite-spin sites. Restricting the Arrhenius analysis exclusively to same-spin hops yields $E_a = 0.18 \pm \red{0.03}$~eV, showing that the activation energy is robust against spin selection.

% =====================================================================
\section*{Discussion}

The picture that emerges from our simulations establishes the oxygen framework as the central actor in small-polaron transport. In \LMPO{} at finite temperature (i.e., in the range 600-900 K)\rev{, we find that 1)} the extra electrons hop between Mn sites a few times per picosecond while the Li ions remain confined to their sites; \rev{2)} \revLi{the polaron visits all Mn sites while the Li ion migrates}\revviii{, including second-shell sites at $\approx$5.5~\AA{} from the Li$^+$ ion (reached already before Li$^+$ migration)}; \rev{3)} freezing the oxygen cage of a single Mn in its $\Mnthree$ geometry excludes that site from hopping without affecting the others; \rev{4)} heavier oxygen slows hopping down through classical framework inertia; \revviii{\red{5) the event-averaged donor--acceptor volume asymmetry $\Delta V$ crosses zero near the electronic switch, consistent with a predominantly breathing reaction coordinate;} and 6) a harmonic estimate from cage-volume fluctuations gives $\lambda_V/4\approx0.20$~eV, comparable to the Arrhenius activation energy of $0.20\pm\red{0.03}$~eV. This agreement supports the relevance of cage elasticity.} All these observations point to the same physical mechanism: thermal fluctuations of the \MnO{} framework transiently bring the donor and acceptor sites into \rev{degeneracy, allowing the electron to transfer, adiabatically in our simulations, and in general with a probability controlled by the diabatic coupling between the two sites~\cite{Marcus1985,Sit2006}.} In this sense, the coordination cage is not a passive spectator that relaxes after charge transfer, but part of the reaction coordinate: the oxidation state of a Mn ion and the geometry of its cage evolve as a single coupled variable. \revviii{Polaron hopping occurs repeatedly while \red{Li$^+$} ions remain in their crystallographic sites, showing that individual local redox transitions do not require simultaneous Li migration. In the \red{$1\times1\times2$} supercell, the polaron also visits Mn sites in the second coordination shell of the \red{Li$^+$} ion (\red{$\approx$5.5~\AA}), the distance at which static calculations locate most of the \red{Li$^+$--polaron} binding, and the \red{Mn$^{2+}$ and Mn$^{3+}$} distance distributions largely overlap at all sites. \red{Therefore}, the electrostatic association does not trap the polaron within the first shell. Long-range dissociation in the dilute limit, however, lies beyond the cell sizes accessible here. The oxygen framework governs the local structural reorganization, whereas \red{Li$^+$} ions shape the average electrostatic landscape and enforce charge neutrality at the mesoscale (i.e., \red{Li$^+$} ions and electrons cannot segregate). Their instantaneous positions, however, do not gate or throttle individual hops.}

\revviii{The activation energy extracted from the dynamics ($E_a=0.20\pm\red{0.03}$~eV) is lower than the barriers computed for Li$^+$ migration in static first-principles calculations (0.25--0.51~eV~\cite{Morgan2004,Shi2016,Sgroi2017,Gao2022}). Although a dynamical and a static barrier are not strictly equivalent, this ordering indicates that the timescale separation observed at 900~K \red{may persist, or even widen, upon cooling if the same local environments, hopping mechanism and barrier ordering persist}. At room temperature, however, Li/vacancy and polaron ordering, or phase separation into LiMnPO$_4$ and a Li-poor phase, as reported experimentally by Chen and Richardson~\cite{Chen2009}, may modify both the migration barriers and the population of local hopping environments. Muon-spin relaxation yields a lower Li self-diffusion barrier ($0.19\pm0.03$~eV~\cite{Sugiyama2020}), but it probes local Li hopping in stoichiometric LiMnPO$_4$ ($x=1$) and is therefore not directly comparable with the partially delithiated compositions simulated here.}

Several physical and methodological aspects place these findings in a broader perspective. \revLi{Our simulations show that} local coordination descriptors (such as octahedral volumes and bond-valence sums) \rev{track the electronic oxidation state to within a few femtoseconds and, through their elasticity, set the energy scale of the hopping barrier}. \revLi{This provides a physical blueprint for redox-aware machine-learning architectures~\cite{Malica2025} which, once equipped with long-range electrostatics, will enable} the exploration of dilute polaron limits, complex defect clustering \revLi{and} low-temperature kinetics. Regarding electron dynamics, \rev{we reiterate that} the OS transitions observed in extended Hubbard FPMD are adiabatic, and the calculated hopping rates correspond to the adiabatic upper limit; non-adiabatic electronic transmission factors~\cite{Marcus1985,Sit2006} would modify the kinetic prefactor, but the fundamental gating by the oxygen cage, the Marcus coincidence condition, and the decoupling from lithium motion are governed by the free-energy landscape and remain unchanged. Likewise, while collinear antiferromagnetic FPMD allows hops between opposite-spin sites, restricting the analysis exclusively to same-spin hops yields an essentially identical activation barrier ($0.18 \pm \red{0.03}$~eV), confirming that the extracted barrier is an intrinsic property of the lattice-gated hopping process.

In conclusion, by following oxidation states in real time with extended Hubbard molecular dynamics, we have shown that small-polaron hopping in olivine \LMPO{} is decoupled from lithium motion and gated by the local oxygen cage: the electron transfers through a geometrically degenerate coincidence state in which donor and acceptor cages are equivalent, at a rate governed by framework inertia and with a barrier of the magnitude set by cage elasticity. \red{From a materials-design perspective, this places} the stiffness and flexibility of the anionic framework, rather than the lithium sublattice, at the centre of electronic transport in polyanionic cathodes, and suggests substitution, strain or pressure as levers to tune it independently of ionic diffusion. Tracking oxidation states while selectively constraining or mass-scaling sublattices offers a general methodology to disentangle ionic, electronic and framework degrees of freedom in mixed ionic--electronic conductors.

% =====================================================================
\section*{Methods}

\subsection*{Extended Hubbard functionals}
\UV{}~\cite{Campo2010} generalizes DFT\,+\,$U$~\cite{Anisimov1991,Dudarev1998} by adding to the DFT energy an extended Hubbard correction,
\begin{equation}
  E_{U+V}=\sum_{I,\sigma}\frac{U^I}{2}\,\mathrm{Tr}\!\left[\mathbf{n}^{II\sigma}\left(1-\mathbf{n}^{II\sigma}\right)\right]-\sum_{I,J,\sigma}^{*}\frac{V^{IJ}}{2}\,\mathrm{Tr}\!\left[\mathbf{n}^{IJ\sigma}\mathbf{n}^{JI\sigma}\right],
\end{equation}
where $I$ and $J$ are atomic sites, $\sigma$ is the spin index, the asterisk indicates that for each $I$ the sum runs over neighbours $J$ up to a given distance, and the generalized occupation matrices $n^{IJ\sigma}_{mm'}=\sum_{\mathbf{k},v}f^\sigma_{\mathbf{k}v}\langle\psi^\sigma_{\mathbf{k}v}|\phi^J_{m'}\rangle\langle\phi^I_m|\psi^\sigma_{\mathbf{k}v}\rangle$ are obtained by projecting the Kohn--Sham states $\psi^\sigma_{\mathbf{k}v}$~\cite{Hohenberg1964,Kohn1965} onto L\"owdin-orthogonalized atomic orbitals $\phi^I_m$~\cite{Lowdin1950,Timrov2020}. The on-site term favours localization on the Mn 3$d$ states, while the inter-site term between Mn 3$d$ and O 2$p$ states accounts for their hybridization~\cite{Campo2010,Cococcioni2019}. As in ref.~\citenum{Malica2025}, we use $U=5.1$~eV and $V=0.7$~eV, obtained by averaging the self-consistent values computed with density-functional perturbation theory~\cite{Timrov2018,Timrov2021,Timrov2022hp} over all Mn sites and Li concentrations; these parameters preserve the digital change of the Mn OSs and yield intercalation voltages in good agreement with experiments~\cite{Malica2025}. Forces are computed via the Hellmann--Feynman theorem, including the contributions of the orthogonalized Hubbard projectors~\cite{Timrov2020}.

\subsection*{Oxidation states}
The OS of each Mn ion is obtained from the eigenvalues of the site-diagonal 3$d$ occupation matrix $\mathbf{n}^{II\sigma}$ (5$\times$5 in each spin channel), following ref.~\citenum{Sit2011}: states with eigenvalues close to 1 are counted as occupied. As a compact descriptor we use the L\"owdin occupation $n=\sum_{i,\sigma}\lambda_i^\sigma$, which equals $\simeq$4.98--5.00 for Mn$^{3+}$ and $\simeq$5.20--5.21 for Mn$^{2+}$ in \UV~\cite{Malica2025}.

\subsection*{First-principles molecular dynamics}
All calculations were performed with the plane-wave pseudopotential implementation of \textsc{Quantum ESPRESSO}~\cite{Giannozzi2009,Giannozzi2017}, using the PBEsol functional~\cite{Perdew2008} and pseudopotentials from the SSSP library~\cite{Bosoni2024} (Mn and Li from GBRV~\cite{Garrity2014}, O and P from Pslibrary~\cite{DalCorso2014}). Kohn--Sham wavefunctions and charge density were expanded up to kinetic-energy cutoffs of 65 and 780~Ry, respectively, with a Gaussian smearing of 0.005~Ry. For the \revviii{25--26}-atom unit cell we used a $\Gamma$-centred $3\times4\times5$ $\mathbf{k}$-point grid; the 49-atom supercell was obtained by doubling the unit cell along $c$, with a correspondingly reduced grid. The antiferromagnetic ground-state configuration was used throughout~\cite{Cococcioni2019}. Born--Oppenheimer FPMD was performed in the NVT ensemble using the stochastic velocity-rescaling thermostat~\cite{Bussi2007} and the Verlet integrator with a time step of 4~fs (0.5~fs for the high-resolution trajectory of Supplementary Figs.~\revviii{S2--S4}), starting from cells and positions relaxed at 0~K. Simulated systems and trajectory lengths are: Li$_{0.5}$MnPO$_4$ unit cell at \rev{600 (32.7~ps), 700 (41.7~ps), 800 (45.1~ps) and 900~K (38.0~ps)}, and at 900~K with $\delta t=0.5$~fs (5.11~ps, 10,216 frames); Li$_{0.25}$MnPO$_4$ unit cell at 900~K (27.8~ps); Li$_{1/8}$MnPO$_4$ supercell at 900~K, unconstrained and with the Mn2O$_6$ cage frozen (13.46~ps each)\red{, and an extension of the unconstrained run to 33.3~ps (Fig.~\ref{fig:association})}.

\paragraph{Constrained dynamics.} In the constrained supercell run, the forces acting on the six O atoms of the Mn2O$_6$ octahedron (Mn--O distances 1.89--2.30~\AA) were set to zero, so that these atoms remain at their initial positions, corresponding to a Mn$^{3+}$ cage; all other atoms, including Mn2, evolve freely.

\paragraph{Fictitious isotope substitution.} The oxygen mass in the equations of motion was set to $m_{\mathrm{O}}=19$, 32 and 79~u, leaving the electronic Hamiltonian unchanged (trajectories of 26.2, 26.8 and \red{26.0}~ps at 900~K, compared with 38.0~ps for $m_{\mathrm{O}}=16$~u).

\subsection*{Hop counting and rates}
A transition on site $i$ is counted when $n_i$ crosses from one basin ($n<5.05$ for Mn$^{3+}$, $n>5.15$ for Mn$^{2+}$) into the other and remains there for at least $\tdw$ (one, two or three time steps, i.e.\ 4, 8 or 12~fs); excursions into the intermediate band $5.05\leq n\leq5.15$ do not change the assigned state. \red{The persistence test is applied only to frames in which all Mn sites of the cell lie within one of the two basins: a frame in which any site is in the intermediate band interrupts the dwell count and leaves the assigned states unchanged.} \revviii{The thresholds correspond to the edges of the two basins at three standard deviations of their thermal fluctuations, $n_{3+}^{\mathrm{mode}}+3\sigma_{3+}$ and $n_{2+}^{\mathrm{mode}}-3\sigma_{2+}$, where $n^{\mathrm{mode}}$ is the peak of the occupation distribution and $\sigma$ is obtained from its outer side (away from the inter-basin tail produced by transfer events), $\sigma\approx0.011$ for both basins. In the unit cells this gives 5.04 and 5.15--5.16. For the Li--Mn distance analysis of the 33.3~ps supercell trajectory (Fig.~\ref{fig:association}), where the Mn$^{2+}$ peak lies slightly lower, the same convention gives $n<5.035$ (Mn$^{3+}$) and $n>5.137$ (Mn$^{2+}$).} This two-threshold hysteresis and the persistence filter suppress recrossings and short-lived unrelaxed excursions. Since each hop changes the OS of exactly two sites, the number of hops is $N_{\mathrm{hop}}=\frac{1}{2}\sum_i N_i$, and the rate per polaron in a cell with $N_{\mathrm{pol}}$ polarons is $\Gamma=N_{\mathrm{hop}}/(N_{\mathrm{pol}}t_{\mathrm{tot}})$; for Li$_{0.5}$MnPO$_4$ ($N_{\mathrm{pol}}=2$, four Mn sites) this is identical to the site average $\langle\Gamma\rangle=\frac{1}{4}\sum_i N_i/t_{\mathrm{tot}}$. Statistical uncertainties are estimated as \red{$\Gamma/\sqrt{N_{\mathrm{hop}}}$ (Poisson statistics of the hops; the two OS changes produced by one hop are not independent events)}. Activation energies are obtained from \rev{Poisson-weighted} least-squares fits of $\ln\langle\Gamma\rangle$ versus $1/T$ \rev{(weighted by} \red{$N_{\mathrm{hop}} = 1/\sigma_{\ln\langle\Gamma\rangle}^2$}\rev{)}.

\subsection*{Structural descriptors and event averaging}
\MnO{} octahedral volumes were computed as the volume of the convex hull of the six O atoms coordinating each Mn~\cite{Barber1996}, under the minimum-image convention. Bond-valence sums were computed as $\mathrm{BVS}=\sum_j\exp[(R_0-d_j)/b]$ over the six Mn--O bonds, with $R_0=1.790$~\AA{} and $b=0.37$~\AA~\cite{Brown1985}. Normalized cross-correlations $C(\tau)$ were computed between the mean-subtracted, variance-normalized time series of each descriptor and of the L\"owdin occupation of the same site. Nearest-neighbour Li--Mn distances were evaluated under the minimum-image convention across all periodic boundaries.

For the hop-aligned coincidence analysis (Fig.~\ref{fig:correlation}c--e), transition times $t_0$ were identified from the mid-point of the L\"owdin occupation crossing ($n=5.10$) in the $\delta t=0.5$~fs trajectory ($N = 16$ hops\red{: the donor--acceptor pairs whose transitions occur within 25~fs of each other, whose occupation difference $n_D-n_A$ crosses zero within $\pm15$~fs of their midpoint, and whose complete $\pm150$~fs window lies inside the 5.11~ps trajectory; with the standard criteria of Supplementary Table~\ref{tab:trajectories}, 20 hops are counted in this trajectory}). Trajectories of donor and acceptor volumes, occupations, and bond-valence sums were aligned at $t_0$ and averaged over a window of $\pm 150$~fs. Single-cage elasticity was extracted from the variance of the octahedral volume $\langle \delta V^2 \rangle$ evaluated strictly within each oxidation-state basin (excluding transition frames), giving $\langle \delta V^2 \rangle = 0.373$~\AA$^6$ for \Mnthree{} and $0.480$~\AA$^6$ for \Mntwo{} (mean: $0.427$~\AA$^6$) in the 900~K high-resolution trajectory, yielding $\lambda_V = 0.81$~eV.

% =====================================================================
\section*{Data availability}
The data supporting the findings of this study will be made available in the Materials Cloud Archive upon publication.

\section*{Code availability}
\UV{} calculations were performed with \textsc{Quantum ESPRESSO} v7.2, open source and available at \url{https://www.quantum-espresso.org}.

\section*{Acknowledgements}
We acknowledge support by the European Commission through the MaX Centre of Excellence for supercomputing applications (grant number 101093374), by the Deutsche Forschungsgemeinschaft (DFG) under Germany's Excellence Strategy (EXC 2077, No.~390741603, University Allowance, University of Bremen) and Lucio Colombi Ciacchi, the host of the ``U Bremen Excellence Chair Program''.

\section*{Competing interests}
The authors declare no competing interests.

% =====================================================================
\bibliographystyle{unsrtnat}
\bibliography{references}

% =====================================================================
\clearpage
\appendix
\setcounter{figure}{0}
\setcounter{table}{0}
\renewcommand{\figurename}{Supplementary Fig.}
\renewcommand{\tablename}{Supplementary Table}
\renewcommand{\thefigure}{S\arabic{figure}}
\renewcommand{\thetable}{S\arabic{table}}

\section*{Supplementary Information}

\subsection*{Summary of First-Principles Molecular Dynamics Trajectories}

Supplementary Table~\ref{tab:trajectories} summarizes all \UV-FPMD trajectories analysed in this study, detailing system composition, temperature, fictitious oxygen mass, integration time step, trajectory duration, number of detected oxidation-state transitions, and resulting hopping rates.

\begin{table}[H]
  \centering
  \footnotesize
  \setlength{\tabcolsep}{3.5pt}
  \caption{\textbf{Overview of \UV-FPMD simulations and polaron hopping statistics.} All runs employ the PBEsol\,+\,$U$\,+\,$V$ functional with $U=5.1$~eV and $V=0.7$~eV. \revviii{Unit cells: 26 atoms (Li$_{0.5}$MnPO$_4$, $a\times b\times c\approx10.15\times6.07\times4.81$~\AA) and 25 atoms (Li$_{0.25}$MnPO$_4$, $\approx9.93\times6.04\times4.84$~\AA); $1\times1\times2$ supercell: 49 atoms ($\approx9.94\times6.03\times9.63$~\AA).} Hop counts and rates correspond to the standard dwell-time threshold $\tdw \geq 12$~fs. \red{Uncertainties are $\Gamma/\sqrt{N_{\mathrm{hop}}}$. The 33.3~ps free supercell run extends the 13.46~ps one (first 2,960 frames shared). For the $\delta t=0.5$~fs run ($\tdw\geq12$~fs = 24 steps), 16 of the hops are used in the event average of Fig.~\ref{fig:correlation}c--e (Methods).}}
  \label{tab:trajectories}
  \vspace{4pt}
  \begin{tabular}{lccccccr}
    \toprule
    Composition & $T$ (K) & $m_{\mathrm{O}}$ (u) & $\delta t$ (fs) & Duration (ps) & Frames & Hops ($N_{\mathrm{hop}}$) & Rate per polaron (ps$^{-1}$) \\
    \midrule
    \multicolumn{8}{l}{\textit{Arrhenius series (unit cell, 2 polarons)}} \\
    Li$_{0.5}$MnPO$_4$ & 600 & 16 (O)  & 4.0 & \rev{32.7} & \rev{8,180}  & \rev{31}  & \rev{$0.47 \pm$}\,\red{$0.09$} \\
    Li$_{0.5}$MnPO$_4$ & 700 & 16 (O)  & 4.0 & \rev{41.7} & \rev{10,414} & \rev{53}  & \rev{$0.64 \pm$}\,\red{$0.09$} \\
    Li$_{0.5}$MnPO$_4$ & 800 & 16 (O)  & 4.0 & 45.1       & \rev{11,266} & \rev{123} & \rev{$1.36 \pm$}\,\red{$0.12$} \\
    Li$_{0.5}$MnPO$_4$ & 900 & 16 (O)  & 4.0 & \rev{38.0} & \rev{9,512}  & \rev{113} & \rev{$1.49 \pm$}\,\red{$0.14$} \\
    \midrule
    \multicolumn{8}{l}{\textit{Oxygen isotope series (unit cell, 2 polarons, $T=900$~K)}} \\
    Li$_{0.5}$MnPO$_4$ & 900 & 16 (O)  & 4.0 & 38.0 & 9,512  & 113 & $1.49 \pm \red{0.14}$ \\
    Li$_{0.5}$MnPO$_4$ & 900 & 19 (F)  & 4.0 & 26.2 & \red{6,556}  & 88  & $1.68 \pm \red{0.18}$ \\
    Li$_{0.5}$MnPO$_4$ & 900 & 32 (S)  & 4.0 & 26.8 & \red{6,698}  & \red{74}  & \red{$1.38 \pm 0.16$} \\
    Li$_{0.5}$MnPO$_4$ & 900 & 79 (Se) & 4.0 & \red{26.0} & \red{6,491}  & \red{56}  & \red{$1.08 \pm 0.14$} \\
    \midrule
    \multicolumn{8}{l}{\textit{High-resolution trajectory (unit cell, 2 polarons)}} \\
    Li$_{0.5}$MnPO$_4$ & 900 & 16 (O)  & 0.5 & 5.11 & 10,216 & \red{20}  & \red{$1.96 \pm 0.44$} \\
    \midrule
    \multicolumn{8}{l}{\textit{Li-dilute trajectory (unit cell, 1 polaron)}} \\
    Li$_{0.25}$MnPO$_4$ & 900 & 16 (O) & 4.0 & 27.8 & 6,948  & \red{133}  & \red{$4.79 \pm 0.42$} \\
    \midrule
    \multicolumn{8}{l}{\textit{Supercell simulations ($1\times1\times2$, 1 polaron, $T=900$~K)}} \\
    Li$_{1/8}$MnPO$_4$ (free)   & 900 & 16 (O) & 4.0 & 13.46 & 3,365 & \red{77} & \red{$5.72 \pm 0.65$} \\
    Li$_{1/8}$MnPO$_4$ (frozen) & 900 & 16 (O) & 4.0 & 13.46 & 3,366 & \red{73} & \red{$5.42 \pm 0.63$} \\
    \revviii{Li$_{1/8}$MnPO$_4$ (free, long)} & \revviii{900} & \revviii{16 (O)} & \revviii{4.0} & \revviii{33.3} & \revviii{8,314} & \red{186} & \red{$5.59 \pm 0.41$} \\
    \bottomrule
  \end{tabular}
\end{table}

\revviii{\subsection*{Li--Mn distance distributions in the unit cell}}
\begin{figure}[H]
  \centering
  \includegraphics[width=\textwidth]{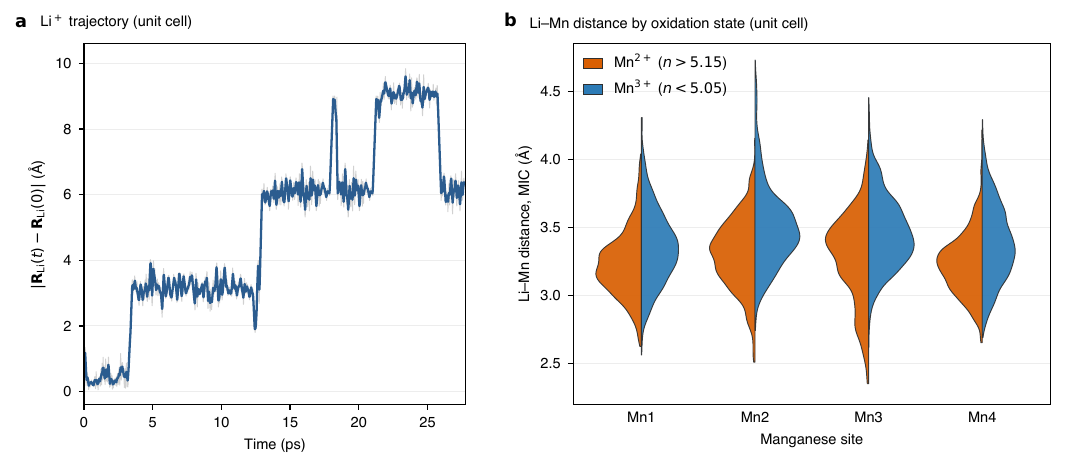}
  \caption{\revviii{\textbf{Li migration and Li--Mn distance distributions in the Li$_{0.25}$MnPO$_4$ unit cell at 900~K.}
  Unit cell (25 atoms: 4 Mn, 4 P, 16 O, 1 Li; $a\times b\times c\approx9.93\times6.04\times4.84$~\AA) with one Li$^+$ ion and one polaron, 27.8~ps (6,948 frames, $\delta t=4$~fs).
  \textbf{a}, Displacement of the Li$^+$ ion, which \red{performs six jumps between nearest-neighbour Li sites ($\approx$3~\AA{} each)}. Faint line: raw data; solid line: 100~fs running average.
  \textbf{b}, Split-violin distributions of the instantaneous Li--Mn distance under the minimum-image convention (MIC, range 2.35--4.73~\AA) for each Mn site, conditioned on the oxidation state of that site: Mn$^{2+}$ (orange, $n>5.15$) and Mn$^{3+}$ (blue, $n<5.05$). All Mn sites lie within the first coordination shell of the Li$^+$ ion, and both oxidation states span the full distance range.}}
  \label{fig:SIunitcell}
\end{figure}

\subsection*{High-temporal-resolution analysis of the electron--lattice coupling}
To resolve the electron--lattice coupling below the 4~fs time step of the production runs, we analysed an independent FPMD trajectory of Li$_{0.5}$MnPO$_4$ at $T=900$~K with $\delta t=0.5$~fs (10,216 frames, 5.11~ps).

\begin{figure}[H]
  \centering
  \includegraphics[width=\textwidth]{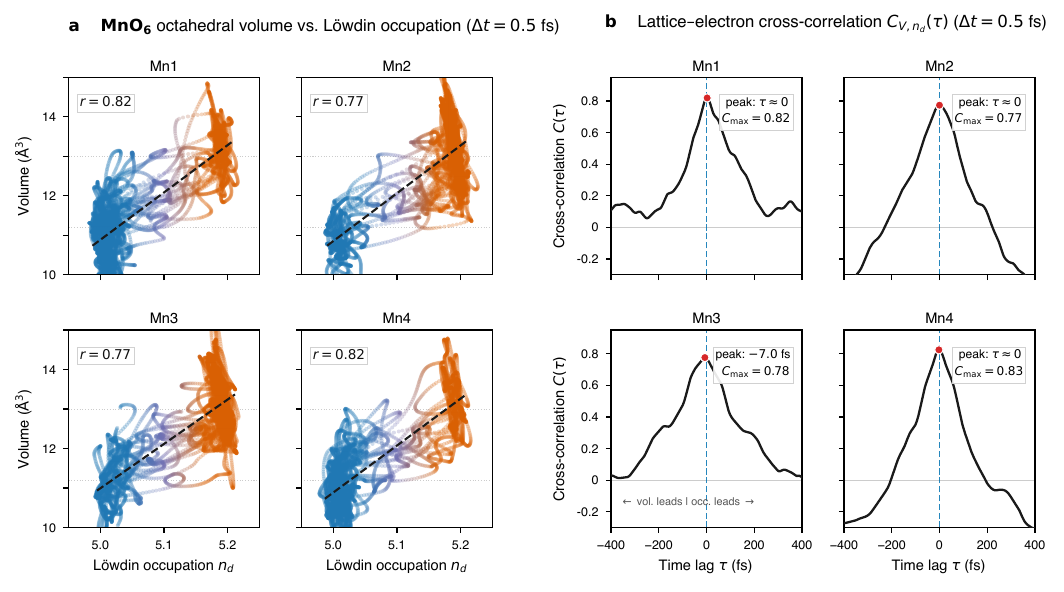}
  \caption{\textbf{Octahedral volume versus L\"owdin occupation at $\delta t=0.5$~fs.} \revviii{Li$_{0.5}$MnPO$_4$ at $T=900$~K, unit cell, 26 atoms: 4 Mn, 4 P, 16 O, 2 Li; $a\times b\times c\approx10.15\times6.07\times4.81$~\AA.} \textbf{a}, Scatter plots for the four Mn sites ($r=0.82$, 0.77, 0.77, 0.82; $\langle r\rangle\simeq0.80$). \textbf{b}, Cross-correlations $C_{V,n}(\tau)$: the maxima lie at $+2.5$, $0.0$, \red{$-7.0$} and \red{$-2.5$}~fs for Mn1--Mn4 (site average \red{$-1.75$}~fs), with $C(0)$ differing from $C_{\max}$ by \red{at most} 0.005. \red{Peak positions are read on the 0.5~fs time grid; a parabolic interpolation shifts them by less than 0.3~fs.}}
\end{figure}

\begin{figure}[H]
  \centering
  \includegraphics[width=\textwidth]{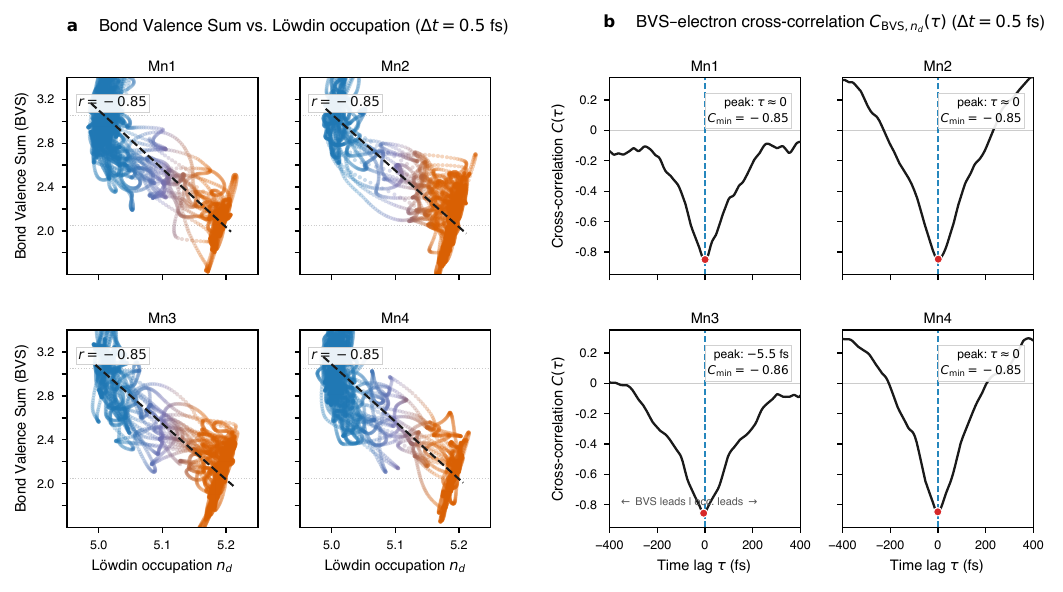}
  \caption{\textbf{Bond-valence sum versus L\"owdin occupation at $\delta t=0.5$~fs.} \revviii{Li$_{0.5}$MnPO$_4$ at $T=900$~K, unit cell, 26 atoms: 4 Mn, 4 P, 16 O, 2 Li; $a\times b\times c\approx10.15\times6.07\times4.81$~\AA.} \textbf{a}, Scatter plots showing a strong anti-correlation ($r=\red{-0.85}$ \red{for all four sites}; $\langle r\rangle\simeq-0.85$). \textbf{b}, Cross-correlations $C_{\mathrm{BVS},n}(\tau)$: the minima lie at \red{$+0.5$, $+1.5$, $-5.5$ and $-0.5$}~fs (site average \red{$-1.0$}~fs), with $C(0)=\red{-0.850}$ versus $C_{\min}=\red{-0.855}$ \red{(Mn3, the largest difference). Peak positions are read on the 0.5~fs time grid; a parabolic interpolation shifts them by less than 0.2~fs.}}
\end{figure}

\begin{figure}[H]
  \centering
  \includegraphics[width=\textwidth]{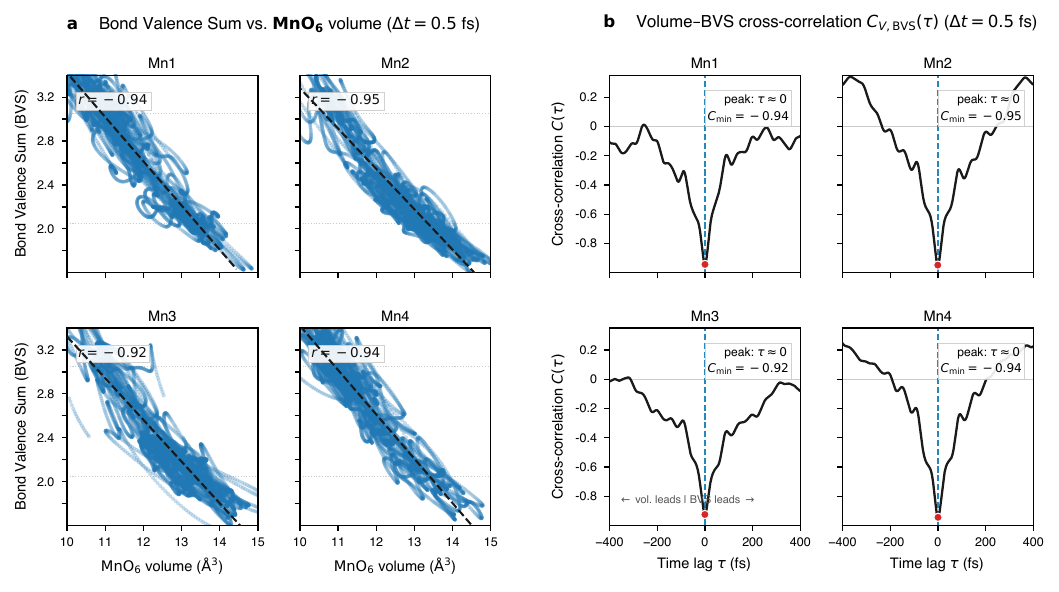}
  \caption{\textbf{Octahedral volume versus bond-valence sum at $\delta t=0.5$~fs.} \revviii{Li$_{0.5}$MnPO$_4$ at $T=900$~K, unit cell, 26 atoms: 4 Mn, 4 P, 16 O, 2 Li; $a\times b\times c\approx10.15\times6.07\times4.81$~\AA.} The two geometric descriptors are strongly anti-correlated ($r=\red{-0.93}$ \red{to} \red{$-0.95$}) and their cross-correlation peaks at $\tau=0$, confirming that they evolve synchronously.}
\end{figure}

\end{document}